\documentclass[twocolumn,showpacs,preprintnumbers,amsmath,amssymb]{revtex4}
\usepackage{epsfig}
\usepackage{graphicx}

\usepackage{epsfig}

\newtheorem{theorem}{Theorem}

\newtheorem{corollary}{Corollary}

\begin{document}


\title{Stabilizer state breeding}

\author{Erik Hostens}
\email{erik.hostens@esat.kuleuven.be}
\affiliation{ESAT-SCD, K.U.Leuven, Kasteelpark Arenberg 10, B-3001 Leuven, Belgium}
\author{Jeroen Dehaene}
\affiliation{ESAT-SCD, K.U.Leuven, Kasteelpark Arenberg 10, B-3001 Leuven, Belgium}
\author{Bart De Moor}
\affiliation{ESAT-SCD, K.U.Leuven, Kasteelpark Arenberg 10, B-3001 Leuven, Belgium}
\date{\today}

\newcommand{\T}{{\cal T}^{(k)}}
\newcommand{\Z}{\mathbb{Z}}
\newcommand{\Gt}{{\cal G}^\perp}
\newcommand{\bra}[1]{\langle #1|}
\newcommand{\ket}[1]{|#1\rangle}
\newcommand{\mat}[2]{\left[\begin{array}{#1} #2 \end{array}\right]}
\newcommand{\arr}[2]{\begin{array}{#1} #2 \end{array}}
\newcommand{\col}[1]{\mathrm{col}\left(#1\right)}
\newcommand{\supp}[1]{\mathrm{supp}\left(#1\right)}

\begin{abstract}
We present a breeding protocol that distills pure copies of any stabilizer state from noisy copies and a pool of predistilled pure copies of the same state, by means of local Clifford operations, Pauli measurements and classical communication.
\end{abstract}


\maketitle

\section{Introduction}
In recent literature, much attention has been paid to the distillation of multipartite entanglement. Distillation is the recovery of entanglement that has been disrupted by the environment, by means of local operations and classical communication. Like quantum error correction codes, entanglement distillation protocols are indispensable procedures towards the realization of many quantum communication and quantum cryptography applications, requiring pure multipartite states that are shared by remote parties \cite{ekert,tele,deutsch,karlsson,hillery,cleve,crep,dur2,hein}.

Many generalizations to a multipartite setting have been found, mainly based on protocols for bipartite entanglement distillation \cite{breeding,hashing}. They can be categorized according to asymptotic (hashing/breeding) \cite{man,asch,lo,CSS,kruspra,krus} versus recurrence-like schemes \cite{murao,acin,man,dur1,asch,lo,kruspra,goyal,krus,glancy,miyake,kay}, to whether they take noise in the recovering operations into account \cite{dur1,asch,kruspra,krus,kay} or to the kind of quantum states they are designed for. The cited references are only suited for Calderbank-Shor-Steane (CSS) states or states that are locally equivalent \footnote{Two states are called locally equivalent if they can be transformed into one another by local operations.} to CSS states (e.g. two-colorable graph states, GHZ states, cluster states), except Ref.~\cite{miyake} (for the W~state) and the recent papers Refs.~\cite{krus,glancy,kay} (for arbitrary stabilizer states).

In this paper, we present a generalization of the breeding protocol for arbitrary stabilizer states, which is to a large extent similar to our previous paper Ref.~\cite{CSS} for CSS states. Again, we define a class of local Clifford operations that distribute the information content of multiple noisy copies of a stabilizer state without destroying the tensor product, i.e. they transform multiple copies of a pure stabilizer state into multiple copies of the same stabilizer state. Breeding, contrary to hashing \cite{CSS}, starts from $k$ noisy copies of an $n$-qubit stabilizer state and a pool of $(1-\gamma)k$ predistilled pure copies of the same state, where $k$ is considered large (asymptotic protocol). After the local Clifford operations, we locally measure the $(1-\gamma)k$ initially pure copies, yielding information on the global state. This information extraction reduces the entropy, resulting in $k$ remaining copies that approach purity (zero entropy) and are suited for the application in mind. The measured copies are afterwards separable and can be discarded. The yield of the protocol is the net output of pure copies that is distilled for every noisy copy, and equals
\[\arr{c}{\frac{\#(\mbox{pure output copies})-\#(\mbox{pure input copies})}{\#(\mbox{noisy input copies})}\\=\frac{k-(1-\gamma)k}{k}=\gamma.}\]

Our result is very much inspired by the work of Glancy, Knill and Vasconcelos \cite{glancy}, in which (nonasymptotically) CSS-H states are purified by any stabilizer code, CSS states by a CSS code and arbitrary stabilizer states by a CSS-H code. Contrary to Ref.~\cite{glancy}, we use a permutation-based approach instead of a code-based approach \cite{eq}. But very similar cases can be distinguished. For CSS states, we use local Clifford operations that are built exclusively of Controlled-NOT (CNOT) operations, which correspond to CSS codes. In Ref.~\cite{CSS} however, more general local Clifford operations are possible for specific CSS states, but CNOTs do always satisfy. Arbitrary local Clifford operations can be used for states that satisfy certain orthogonality conditions \cite{CSS}. It can be verified that these states are CSS-H states. And now we will show that local Clifford operations, built of CNOTs and represented by an orthogonal matrix, corresponding to CSS-H codes, are suited for the distillation of arbitrary stabilizer states.

This paper is organized as follows. In Sec.~\ref{sectionbinary}, we give a brief overview of the binary linear algebra framework in which we describe the stabilizer formalism. In the past, this ``binary picture'' has been proved useful in the context of distillation protocols \cite{DVD:03,eq,CSS,hos}. In Sec.~\ref{sectioninfo}, we summarize some properties of the strongly typical set that we will need to calculate $\gamma$. For the sake of readability, we prefer to make proofs not as rigorous as in Ref.~\cite{CSS}, since they are highly similar. In Sec.~\ref{sectionbreeding}, we explain the breeding protocol and calculate $\gamma$. We illustrate this with a typical example in Sec.~\ref{sectionex}. We conclude in Sec.~\ref{sectionconclusion}.

\section{Preliminaries}

\subsection{The stabilizer formalism in the binary picture}\label{sectionbinary}

In this section, we give a brief overview of the binary matrix description of Pauli operations, stabilizer states and Clifford operations. For a more elaborate discussion, we refer to Refs.~\cite{D:03, CSS}. In the following, all addition and multiplication of binary objects is performed modulo 2.

\subsubsection{Pauli operations}
We use the following notation for Pauli matrices.
\[\arr{ccccl}{\sigma_{00} &=& I_2 &=& \mat{rr}{1&0\\0&1},\\
\sigma_{01} &=& \sigma_x &=& \mat{rr}{0&1\\1&0},\\
\sigma_{10} &=& \sigma_z &=& \mat{rr}{1&0\\0&-1},\\
\sigma_{11} &=& \sigma_y &=& \mat{rr}{0&-i\\i&0}.
} \]
Let $v,w\in\Z_2^n$ and $a=\mat{cc}{v\\w}$, then we denote
\begin{equation*}
\sigma_a=\sigma_{v_1w_1}\otimes\ldots\otimes\sigma_{v_nw_n}.
\end{equation*}
The Pauli group on $n$ qubits is defined to contain all tensor products $\sigma_a$ of Pauli matrices with an additional complex phase factor in $\{1,i,-1,-i\}$. In this paper, we only consider Hermitian Pauli operations, so we may exclude imaginary phase factors. It can be verified that Pauli operations satisfy the following commutation relation:
\begin{equation}\label{comm}
\sigma_a\sigma_b=(-1)^{a^TPb}\sigma_b\sigma_a,~\mbox{where}~P=\mat{cc}{0&I_n\\I_n&0}.
\end{equation}

\subsubsection{Stabilizer states}
A stabilizer state on $n$ qubits is the simultaneous eigenvector, with eigenvalues 1, of $n$ commuting Hermitian Pauli operations $(-1)^{b_i}\sigma_{s_i}$, where $s_i\in\Z_2^{2n}$ are linearly independent and $b_i\in\Z_2$, for $i=1,\ldots, n$. These Pauli operations generate an Abelian subgroup of the Pauli group on $n$ qubits, called the stabilizer $\cal S$. We will assemble the vectors $s_i$ as the columns of a matrix $S\in\Z_2^{2n\times n}$ and the bits $b_i$ in a vector $b\in\Z_2^n$. With (\ref{comm}), commutativity of the stabilizer is reflected by $S^TPS=0$. The representation of $\cal S$ by $S$ and $b$ is not unique, as every other generating set of $\cal S$ yields an equivalent description. In the binary picture, a change from one generating set to another is represented by an invertible linear transformation $R\in\Z_2^{n\times n}$ acting on the right on $S$ and acting appropriately on $b$. We have
\begin{equation}\label{R}\arr{ccl}{
S' &=& SR \\
b' &=& R^Tb+d
}\end{equation}
where $d\in\Z_2^n$ is a function of $S$ and $R$ but not of $b$ \cite{D:03}. In the context of distillation protocols, $d$ can always be made zero \cite{CSS}.

Each $S$ defines a total of $2^n$ orthogonal stabilizer states, one for each $b\in\Z_2^n$. As a consequence, all stabilizer states defined by $S$ constitute a basis for ${\cal H}^{\otimes n}$, where $\cal H$ is the Hilbert space of one qubit. In the following, we will refer to this basis as the $S$-basis.

Let $\ket{\psi_1}$ and $\ket{\psi_2}$ be two stabilizer states represented by $S_1=\mat{c}{S_{1(z)}\\S_{1(x)}}, b_1$ and $S_2=\mat{c}{S_{2(z)}\\S_{2(x)}}, b_2$ respectively. Then $\ket{\psi_1}\otimes\ket{\psi_2}$ is a stabilizer state represented by
\begin{equation}\label{stabprod}
\mat{cc}{S_{1(z)} & 0\\0 & S_{2(z)}\\S_{1(x)} & 0\\0 & S_{2(x)}},~\mat{cc}{b_1\\b_2}.\end{equation}

If the phase factor vector $b$ of a stabilizer state $\ket{\psi}$ represented by $S=\mat{c}{S_z\\S_x}$ is unknown, information on $b$ by local measurements can be extracted as follows. From the definition of $\ket{\psi}$, it follows that $\sigma_{Sv}\ket{\psi}=(-1)^{v^Tb+d}\ket{\psi}$ for every $v\in\Z_2^n$, where $d\in\Z_2$ is a function of $S$ and $v$ \cite{D:03}. Consequently, we can find the inner product $v^Tb$ by determining the eigenvalue of $\ket{\psi}$ for $\sigma_{Sv}$. We determine the eigenvalue for $\sigma_g$ by locally measuring its nontrivial factors $\sigma_{g_ig_{n+i}}$ on qubits $i=1,\ldots,n,$ and taking the product of the outcomes. It follows that the eigenvalue for $\sigma_g$ and $\sigma_h$ can be determined both if and only if the factors $\sigma_{g_ig_{n+i}}$ and $\sigma_{h_ih_{n+i}}$ commute, for all $i$. In the binary picture: let $\cal M$ be a partition of $\{1,\ldots,n\}$ into three disjunct subsets $M_z$, $M_x$ and $M_y$, and perform a $\sigma_z$, $\sigma_x$ or $\sigma_y$ measurement on qubits $i\in M_z$, $M_x$ or $M_y$ respectively. Note that there exists $3^n$ such partitions. Then from the outcomes of these measurements we can calculate $v^Tb$ for all $v$ that satisfy
\begin{equation}\label{meascond}\arr{ccl}{
\supp{S_zv}\backslash\supp{S_xv} &\subset& M_z,\\
\supp{S_xv}\backslash\supp{S_zv} &\subset& M_x,\\
\supp{S_zv}\cap\supp{S_xv} &\subset& M_y,}\end{equation}
where the \emph{support} of a vector $v\in\Z_2^n$ is defined as
\[\supp{v}=\{i\in\{1,\ldots,n\}~|~v_i=1\}.\]
All $v$ satisfying (\ref{meascond}) constitute a subspace ${\cal V}({\cal M})$ of $\Z_2^n$. We denote $\mathrm{dim}[{\cal V}({\cal M})]$ by $n({\cal M})$.

\subsubsection{Clifford operations}
A Clifford operation $Q$, by definition, maps the Pauli group to itself under conjugation:
\[Q\sigma_a Q^{\dag}=(-1)^{\delta}\sigma_b.\]
In the binary picture, a Clifford operation is represented by a matrix $C\in\Z_2^{2n\times 2n}$ and a vector $h\in\Z_2^{2n}$, where $C$ is symplectic or $C^TPC=P$ \cite{D:03}. The image of a Pauli operation $\sigma_{a}$ under the action of a Clifford operation is then given by $(-1)^{\epsilon}\sigma_{Ca}$, where $\epsilon$ is function of $C,h$ and $a$.

If a stabilizer state $\ket{\psi}$, represented by $S$ and $b$, is operated on by a Clifford operation $Q$, represented by $C$ and $h$, $Q\ket{\psi}$ is a new stabilizer state whose stabilizer is given by $Q{\cal S}Q^{\dag}$. As a result, this stabilizer is represented by
\begin{equation}\label{CS}\arr{ccl}{
S' &=& CS \\
b' &=& b+f
}\end{equation}
where $f$ is independent of $b$ and can always be made zero, by performing an extra Pauli operation $\sigma_g$ before  the Clifford operation, where $S^TPg=f$ \cite{CSS}.

Let $Q_1$ and $Q_2$ be two Clifford operations represented by $\mat{cc}{A_1&B_1\\C_1&D_1}$ and $\mat{cc}{A_2&B_2\\C_2&D_2}$ respectively, where all blocks are in $\Z_2^{n\times n}$. Then $Q_1\otimes Q_2$ is a Clifford operation represented by
\begin{equation}\label{clifprod}
\mat{cccc}{A_1 & 0 & B_1 & 0\\0 & A_2 & 0 & B_2\\C_1 & 0 & D_1 & 0\\0 & C_2 & 0 & D_2}.\end{equation}

The breeding protocol that will be introduced in Sec.~\ref{sectionbreeding}, considers $\bar{k}$ stabilizer states on $n$ qubits that are shared by $n$ remote parties, each holding corresponding qubits of all $\bar{k}$ states. Each stabilizer state is represented by the same $S$. According to (\ref{stabprod}), the overall state is then represented by
\begin{equation*}
\mat{c}{I_{\bar k}\otimes S_z \\I_{\bar k}\otimes S_x},~\bar{b}'=\mat{c}{b_1\\\vdots\\b_{\bar k}}.
\end{equation*}
Since it is more convenient to arrange all qubits per party, we rewrite the stabilizer matrix by permuting rows and columns as follows:
\begin{equation}\label{copies}
\mat{c}{S_z\otimes I_{\bar k} \\ S_x\otimes I_{\bar k}}=S\otimes I_{\bar k},~\bar{b}=\mat{c}{\bar{b}_1\\\vdots\\\bar{b}_n},
\end{equation}
where the entries of $\bar{b}'$ are permuted appropriately into $\bar{b}\in\Z_2^{n\bar{k}}$. All parties perform the same local Clifford operation, built only of CNOTs. It is shown in Ref.~\cite{D:03} that a Clifford operation, composed of CNOTs, is represented by
\begin{equation*}
C=\left[\begin{array}{cc}A&0\\0&A^{-T}\end{array}\right],
\end{equation*}
where $A$ is invertible. With (\ref{clifprod}), the overall Clifford operation is then represented by
\begin{equation}\label{LC}
\mat{cc}{I_n\otimes A & 0\\ 0 & I_n\otimes A^{-T} },\end{equation}
Furthermore, we will demand that $A\in\Z_2^{\bar{k}\times\bar{k}}$ is orthogonal, i.e. $A^{-T}=A$. In that case, the representation of the overall state after the local Clifford operation can be transformed back into the original form of (\ref{copies}) by multiplication with $R=I_n\otimes A^T$ on the right, since
\begin{equation*}
(I_{2n}\otimes A)(S \otimes I_{\bar k})(I_n\otimes A^T) =  S \otimes I_{\bar k}.
\end{equation*}
Using (\ref{R}) and (\ref{CS}), $\bar{b}$ is then transformed as follows:
\begin{equation}\label{permutation}
\bar{b}\rightarrow (I_n\otimes A)\bar{b}.
\end{equation}

\subsection{Strongly typical set}\label{sectioninfo}
In this section, we introduce the information-theoretical notion of a \emph{strongly typical set}. For proofs, we refer to Ref.~\cite{CSS}. Good introductory material on information theory can be found in Ref.~\cite{CT}. We use the compact notation $x\approx y$ for $x=y\pm\epsilon$, where $\epsilon\rightarrow 0$ for $k\rightarrow\infty$.

Let $X=(X_1,\ldots,X_k)$ be a sequence of independent and identically distributed discrete random variables,
each having event set $\Omega$ with probability function $p: \Omega\rightarrow [0,1]: a\rightarrow p(a)$. The
strongly typical set $\T_\epsilon$ is defined as the set of sequences $x=(x_1,\ldots,x_k)\in\Omega^k$ for which the
sample frequencies $f_a(x)=|\{x_i~|~x_i=a\}|/k$ are close to the true values $p(a)$, or:
\[x\in\T_\epsilon~\Leftrightarrow~|f_a(x)-p(a)|<\epsilon,~\forall a\in\Omega.\]
It can be verified that $p(\T_\epsilon)\geq 1-\delta$, where $\delta=O(k^{-1}\epsilon^{-2})$. Therefore, $p(\T_\epsilon)\approx 1$ for $k\rightarrow\infty$. In words, a random sequence $x$ will almost certainly be contained in the strongly typical set.

Let $\Omega$ be partitioned into subsets $\Omega_j$ ($j=1,\ldots, t$). We define the function
\[y(x)=(\Omega_{j_1},\ldots,\Omega_{j_k}),~\mbox{where}~x_i\in\Omega_{j_i},~\mbox{for}~i=1,\ldots, k.\]
In Sec.~\ref{sectionmeas}, we will encounter the following problem. Given some $u\in\T_\epsilon$, calculate the number $|{\cal N}_u|$ of sequences $v\in\T_\epsilon$ that satisfy $y(v)=y(u)$, or
\[{\cal N}_u=\{v\in\T_\epsilon~|~y(v)=y(u)\}.\]
It can be verified that
\[\frac{\log_2|{\cal N}_u|}{k}\approx H(X)-H(Y),\]
where
\[\arr{rrcl}{
 & H(X) &=& -\sum\limits_{a\in\Omega} p(a)\log_2 p(a)\\
\mbox{and} & H(Y) &=& -\sum\limits_{j=1}^t p(\Omega_j)\log_2 p(\Omega_j) }\] are the entropies of $X$ and $y(X)$
respectively.

\section{Breeding}\label{sectionbreeding}
In this section, we show how the protocol works and calculate the yield. The standard information-theoretical interpretation of hashing/breeding protocols is as follows. We have $k$ copies of a mixed stabilizer state $\rho$, that we wish to purify. The global state of the copies can be regarded as a classical ensemble of pure states or, equivalently, as an unknown pure state of which we know the a-priori probabilities. With vanishing error probability, this unknown state can then be assumed to be contained in the strongly typical set $\T$. The protocol consists of local Clifford operations, followed by local measurements. These measurements yield information on the unknown state, eliminating all elements $\T$ that do not match the outcomes. The protocol ends when all elements but one are eliminated from $\T$ and we are left with a known pure state.

This section is organized as follows. In Sec.~\ref{sectionprotocol}, we go into more detail on the information-theoretical interpretation as explained above. In Sec.~\ref{sectionextract}, we show which measurements maximize the amount of extracted information or, equivalently, the probability that an element of $\T$ is eliminated. Finally, in Sec.~\ref{sectionmeas}, we calculate the minimal number of measurements necessary to reduce $\T$ to a singleton with probability approaching unity. The remaining element is the formerly unknown pure state.

\subsection{Protocol}\label{sectionprotocol}
As noted in Sec.~\ref{sectionbinary}, all $2^n$ stabilizer states represented by the same $S\in\Z_2^{2n\times n}$ constitute a basis for ${\cal H}^{\otimes n}$, which we call the $S$-basis. The protocol starts with $k$ identical copies of a mixed state $\rho$ that is diagonal in this basis and $(1-\gamma)k$ copies of the pure stabilizer state represented by $S$ and $b=0$. The copies of $\rho$ could for instance result from distributing $k$ pure copies via imperfect quantum channels. If $\rho$ is not diagonal in the $S$-basis, it can always be made that way by performing a local POVM measurement \cite{asch}. We have
\begin{equation*}
\rho=\sum\limits_{b\in\Z_2^n}p(b)\ket{\psi_b}\bra{\psi_b},
\end{equation*}
where $\ket{\psi_b}$ stands for the stabilizer state represented by $S$ and $b$. The mixed state $\rho$ can be regarded as a statistical ensemble of pure states $\ket{\psi_b}$ with probabilities $p(b)$. Consequently, $\rho^{\otimes k}$ is an ensemble of pure states represented by $S\otimes I_k$ and $\tilde{b}$, with probabilities
\begin{equation*}
p(\tilde{b})=p(\tilde{b}')=\prod_{i=1}^{k} p(b_i).
\end{equation*}
Again, the entries of $\tilde{b}$ correspond to the $nk$ phase factors ordered per party instead of per copy like $\tilde{b}'$. We define $\bar{k}=(2-\gamma)k$ and $\bar{b}$ from $\tilde{b}$ as follows:
\[\bar{b}_i=\mat{c}{\tilde{b}_i\\0},~\mbox{for}~i=1,\ldots,n,\]
and we have that $\rho^{\otimes k}\otimes (\ket{\psi_0}\bra{\psi_0})^{\otimes (1-\gamma)k}$ is represented by (\ref{copies}).

The ensemble can be interpreted as an unknown pure state. The probability that this state is represented by $\tilde{b}$ is then equal to $p(\tilde{b})$. Let this unknown pure state be represented by $\tilde{u}$. With probability approaching unity, $\tilde{u}$ is contained in the set $\T$, defined as in Sec.~\ref{sectioninfo}. Here, $\Omega$ is the set of all $b\in\Z_2^n$. So with negligible error probability, we may assume that $\tilde{u}\in\T$.

The protocol consists of the following steps:
\begin{enumerate}
\item Each party applies local Clifford operations (\ref{LC}) with orthogonal $A$, that, with (\ref{permutation}), result in the transformation $\bar{u}_i\rightarrow A\bar{u}_i$, for $i=1,\ldots,n$.
\item The resulting last $(1-\gamma)k$ copies (which were the initial pure states) are measured locally, yielding information on $\tilde{u}$. Afterwards, these copies are in a separable state and can be discarded.
\end{enumerate}
After each measurement, we eliminate every $\tilde{b}\in\T$ that is inconsistent with the measurement outcome. The protocol ends when all $\tilde{b}\neq\tilde{u}$ are eliminated from $\T$ and only $\tilde{u}$ is left.

\subsection{Maximal information extraction}\label{sectionextract}
The local Clifford operations (\ref{LC}) transform each $\bar{b_i}$ to $A\bar{b_i}$, for $i=1,\ldots,n$. As the
last $(1-\gamma)k$ entries of $\bar{b_i}$ are zero, and the last $(1-\gamma)k$ copies are the ones measured, the
only relevant part of $A$ is the lower left $(1-\gamma)k\times k$ part. We define $Q$ as the transpose of this
part. We show in Appendix~\ref{apportho}, for arbitrary full rank $Q$, how to construct an orthogonal matrix $A$
with lower left $(1-\gamma)k\times k$ part ${Q'}^T$, where $Q'$ is either equal to $Q$ or equivalent to
$Q$ for the protocol.

We calculate the probability that some $\tilde{b}\neq\tilde{u}$ is not eliminated after the local measurement of one of the $(1-\gamma)k$ ancillary $n$-qubit states (take the $i$th). Let $\cal M$ be the partition according to which this state is measured and $q$ the $i$th column of $Q$ . If we organize $\tilde{b}$ in the following matrix:
\[\tilde{B}=\mat{ccc}{b_1 & \cdots & b_k}=\mat{c}{\tilde{b}_1^T \\ \vdots \\ \tilde{b}_n^T},\]
then, by the local Clifford operations, $b_{k+i}$ is transformed into $\tilde{B}q$. Defining $\tilde{U}$ in the same way, $u_{k+i}$ is transformed into $\tilde{U}q$. We know from Sec.~\ref{sectionbinary} that the measurement reveals
$V^Tu_{k+i}=V^T\tilde{U}q$, where $\col{V}$, the column space of $V\in\Z_2^{n\times n({\cal M})}$, equals
${\cal V}({\cal M})$. The information contained by $V^T\tilde{U}q$ is maximal when $q$ is uniformly distributed
over $\Z_2^{k}$. Indeed, as $V^T\tilde{U}q$ is a linear function of $q$, then $V^T\tilde{U}q$ will be
distributed uniformly too (over the range of $V^T\tilde{U}$).

If and only if $V^T(\tilde{B}+\tilde{U})q=0$, then $\tilde{b}$ is not eliminated from $\T$ by the measurement. Indeed, in that case $\tilde{b}$ would have the same measurement outcome $V^T\tilde{B}q$ as $\tilde{u}$. Let $\Delta\tilde{b}=\tilde{b}+\tilde{u}$ and $d({\cal M},\Delta\tilde{b})$ the rank of $V^T\Delta\tilde{B}$,
which is at most $n({\cal M})$. Then the probability of $V^T\Delta\tilde{B}q=0$ is equal to $2^{-d({\cal
M},\Delta\tilde{b})}$, as this is the inverse of the number of elements in the range of $V^T\Delta\tilde{B}$. The same reasoning can be applied to all measurements. Consequently, the probability that
some $\tilde{b}$ will not be eliminated after all measurements is equal to
\begin{equation}\label{survprob}
2^{-k\sum\limits_{\cal M}m({\cal M})d({\cal M},\Delta\tilde{b})},
\end{equation}
where $km({\cal M})$ is the number of measurements corresponding to partition $\cal M$. Note that $\sum_{\cal M} m({\cal M})=1-\gamma$.

\subsection{Minimal number of measurements}\label{sectionmeas}

So far we have given an information-theoretical interpretation of the protocol: we start with an unknown pure state (represented by $\tilde{u}$), which is almost certainly contained in $\T$. Consecutive measurements rule out all inconsistent $\tilde{b}\in\T$. The probability that a particular $\tilde{b}\neq\tilde{u}$ survives this process equals (\ref{survprob}). Consequently, the probability that any $\tilde{b}\neq\tilde{u}$ survives the process is equal to
\begin{equation}\label{ps}
\sum_{f\not\equiv 0} N_f^\ast 2^{-k\sum\limits_{\cal M}m({\cal M})f({\cal M})}
\end{equation}
where the sum runs over all functions $f: f({\cal M})\in\{0,1,\ldots,n({\cal M})\}$ that are not identical to zero.
$N_f^\ast$ is the number of $\tilde{b}\in\T$ for which $d({\cal
M},\Delta\tilde{b})=\mathrm{dim}[\Delta\tilde{B}^T{\cal V}({\cal M})]=f({\cal M})$. Let $N_f^\ast =
2^{k\alpha_f^\ast}$. Then (\ref{ps}) vanishes if and only if the following inequalities hold:
\begin{equation}\label{ast}
\sum\limits_{\cal M}m({\cal M})f({\cal M}) > \alpha_f^\ast,~\mbox{for all}~f\not\equiv 0.
\end{equation}
Let $N_f = 2^{k\alpha_f}$ be the number of $\tilde{b}\in\T$ for which $d({\cal M},\Delta\tilde{b})\leq
f({\cal M})$. Evidently,
\begin{equation}\label{sumN}
N_f=\sum_{f'\leq f}N_{f'}^\ast,
\end{equation}
where $f'\leq f$ stands for: $f'({\cal M})\leq f({\cal M})$ for all $\cal M$. The inequalities
\begin{equation}\label{noast}
\sum\limits_{\cal M}m({\cal M})f({\cal M}) > \alpha_f,~\mbox{for all}~f\not\equiv 0,
\end{equation}
are equivalent to (\ref{ast}). Indeed, it follows from (\ref{sumN}) that $\alpha_f\approx\alpha_{f'}\approx\alpha_{f'}^\ast$ for some $f'\leq f$ \footnote{We make use of the fact that if $2^{k\alpha}=\sum\limits_{i=1}^g 2^{k\alpha_i}$ for some natural number $g$ (independent of $k$) and $k\rightarrow\infty$, then $\alpha\approx\alpha_i$ for some $i$.}. Since $\sum\limits_{\cal M}m({\cal M})f'({\cal M}) > \alpha_{f'}^\ast\approx\alpha_{f'}\approx\alpha_f$ implies $\sum\limits_{\cal M}m({\cal M})f({\cal M}) > \alpha_{f}$, a solution to (\ref{noast}) is also a solution to (\ref{ast}) and vice versa.

This leaves us to calculate $N_f$. Let ${\cal G}_f({\cal M})$ be a $n({\cal M})-f({\cal M})$ dimensional
subspace of ${\cal V}({\cal M})$. For a space $\cal G$, we define
\[{\cal L}({\cal G})=\{\Delta\tilde{b}\in\Z_2^{nk}~|~G^T\Delta\tilde{B}=0,~\mbox{where}~\col{G}={\cal G}\}.\]
It follows that $d({\cal M},\Delta\tilde{b})\leq f({\cal M})$ for all $\Delta\tilde{b}\in{\cal L}[{\cal
G}_f({\cal M})]$. We then have
\[N_f = |\bigcup_{{\cal G}_f}{\cal L}\left(\sum\limits_{\cal M}{\cal G}_f({\cal M})\right) \cap \T|\]
where the union runs through all functions ${\cal G}_f:{\cal M}\rightarrow {\cal G}_f({\cal M})$, where every ${\cal
G}_f({\cal M})$ is a subspace of ${\cal V}({\cal M})$ with dimension $n({\cal M})-f({\cal M})$. It follows that
\[N_f = r\max_{{\cal G}_f}|{\cal L}\left(\sum\limits_{\cal M}{\cal G}_f({\cal M})\right) \cap \T|,\]
where $1\leq r\leq$ the total number of functions ${\cal G}_f$, which is independent of $k$. Therefore,
\[\alpha_f \approx \log_2(\max_{{\cal G}_f}|{\cal L}\left(\sum\limits_{\cal M}{\cal G}_f({\cal M})\right) \cap \T|)/k.\]

We now calculate $\log_2|{\cal L}({\cal G}) \cap \T|/k$. Note that $\Delta\tilde{b}\in {\cal L}({\cal G})$ if and only
if $G^T\Delta b_i=0$, for $i=1,\ldots, k$. The cosets $\Omega_j$ ($j=1,\ldots, t$) of the space
$\Gt=\{v\in\Z_2^n|G^Tv=0\}$ constitute a partition of $\Z_2^n$. Note that $t=2^{\mathrm{dim}({\cal G})}$. We want to know the logarithm, divided by $k$, of the number of $\tilde{b}\in\T$
for which $b_i$ is in the same coset as $u_i$, for all $i=1,\ldots, k$. We know from Sec.~\ref{sectioninfo} that
this is approximately
\[\begin{array}[t]{ccccl}
H-C({\cal G}) & \mbox{where} & H &=& -\sum\limits_{b\in\Z_2^n}p(b)\log_2 p(b)\\
& \mbox{and} & C({\cal G}) &=& -\sum\limits_{j=1}^t p(\Omega_j)\log_2 p(\Omega_j).
\end{array}\]
Let 
\[H_f=\min\limits_{{\cal G}_f} C\left(\sum\limits_{\cal M}{\cal G}_f({\cal M})\right).\]
It follows that
\begin{equation}\label{alpha}
\alpha_f\approx H-H_f.
\end{equation}

The yield $\gamma$ is maximized by minimizing the total number of measurements. With (\ref{noast}) and
(\ref{alpha}), this results in the following linear programming (LP) problem:
\[\arr{cl}{
\mbox{minimize} & \sum\limits_{\cal M} m({\cal M}) \\[4mm]
\mbox{subject to} &  \sum\limits_{\cal M}m({\cal M})f({\cal M}) > H-H_f,~\mbox{for all}~f\not\equiv 0. }\]

\section{Illustration with a three-colorable graph state}\label{sectionex}
\begin{figure}
\includegraphics[width=0.45\textwidth]{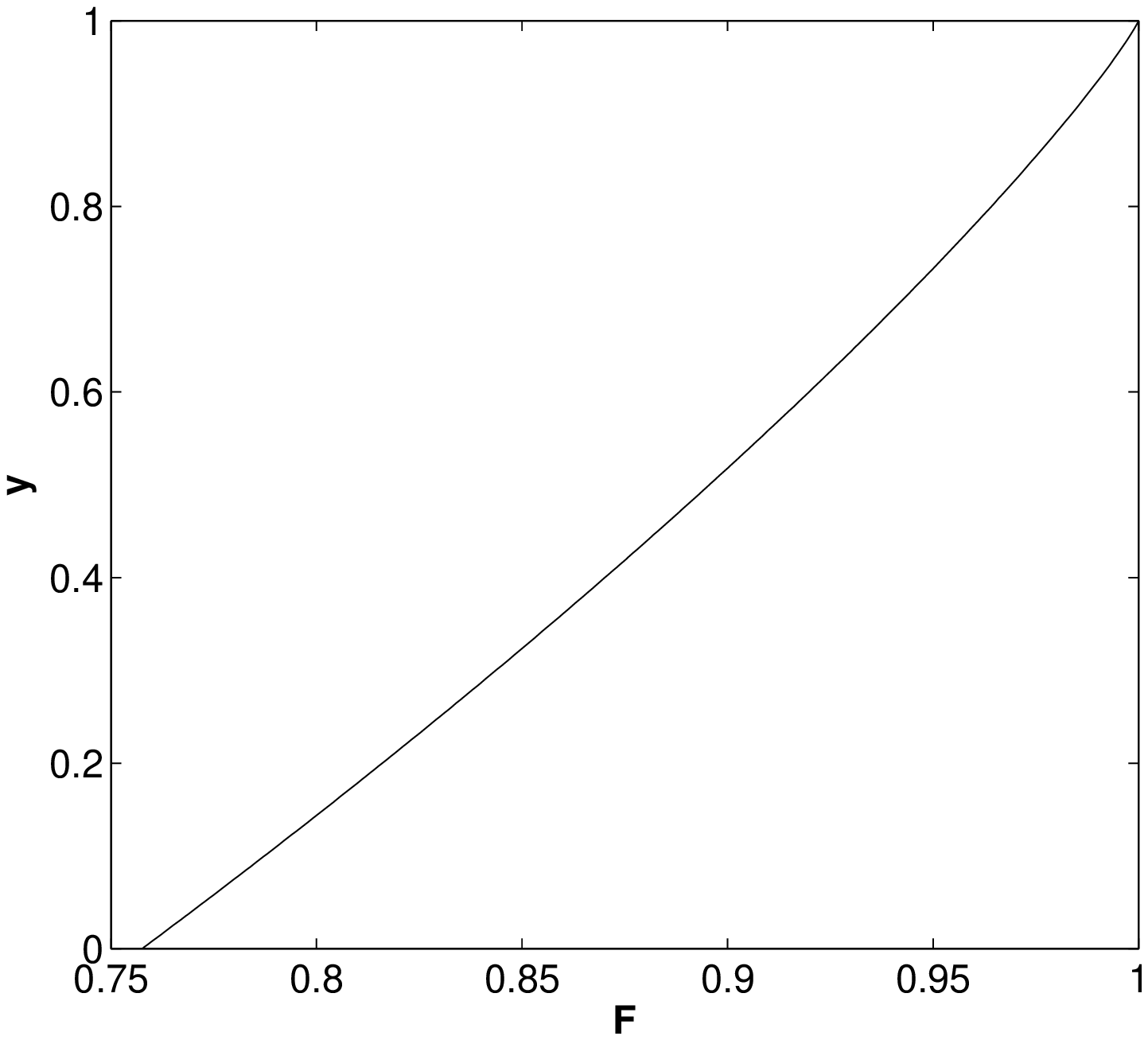}
\caption{\label{figgamma} the yield $\gamma$ of the protocol for the state $\rho=F\ket{\psi_0}\bra{\psi_0}+(1-F)/31(\openone - \ket{\psi_0}\bra{\psi_0})$ as a function of $F$.}
\end{figure}

We illustrate our protocol with an example. Any stabilizer state is locally equivalent to a graph state, i.e. any stabilizer state can be reversibly transformed into a graph state by means of one-qubit Clifford operations \cite{schlingi,LU}. In order not to be covered by our previous result \cite{CSS}, the given example should be a state that is not equivalent to a CSS state or a two-colorable graph state. The 5-qubit ring state, which is an example of a three-colorable graph state \cite{krus}, is such a state. We calculate the yield for the following mixture:
\[\rho=F\ket{\psi_0}\bra{\psi_0}+\frac{1-F}{2^5-1}\sum\limits_{b\in\Z_2^5\backslash\{0\}}\ket{\psi_b}\bra{\psi_b},\]
where $\ket{\psi_b}$ is binary represented by $S$ and $b$, and
\[S=\mat{c}{\theta\\I_5},~\mbox{with}~\theta=\mat{ccccc}{0&0&1&1&0\\0&0&0&1&1\\1&0&0&0&1\\1&1&0&0&0\\0&1&1&0&0}.\]
We will only consider the following partitions:
\[\arr{cccc}{
{\cal M}_1: & M_z=\{3,4,5\}, & M_x=\{1,2\}, & M_y=\emptyset;\\
{\cal M}_2: & M_z=\{1,4,5\}, & M_x=\{2,3\}, & M_y=\emptyset;\\
{\cal M}_3: & M_z=\{1,2,5\}, & M_x=\{3,4\}, & M_y=\emptyset;\\
{\cal M}_4: & M_z=\{1,2,3\}, & M_x=\{4,5\}, & M_y=\emptyset;\\
{\cal M}_5: & M_z=\{2,3,4\}, & M_x=\{1,5\}, & M_y=\emptyset.}\]
By restricting to these partitions, we risk finding a suboptimal solution. However, we will show below that the solution found is in fact optimal. It can be verified that these partitions satisfy $n({\cal M}_i)=2$, and for no other partition $n({\cal M})>2$. For symmetry reasons, we may assume that the optimal $m({\cal M}_i)$ will be equal for all ${\cal M}_i$. We have
\[{\cal V}({\cal M}_1)=\col{V}~\mbox{where}~V=\mat{cc}{1&0\\0&1\\0&0\\0&0\\0&0},\]
and $f({\cal M}_1)$ can be either 0, 1 or 2:
\begin{itemize}
\item[$\rightarrow$] if $f({\cal M}_1)=0$ then ${\cal G}_f({\cal M}_1)={\cal V}({\cal M}_1)$,
\item[$\rightarrow$] if $f({\cal M}_1)=1$ then ${\cal G}_f({\cal M}_1)=\col{e_1}~\mbox{or}~\col{e_2}$,
\item[$\rightarrow$] if $f({\cal M}_1)=2$ then ${\cal G}_f({\cal M}_1)=\{0\}$,
\end{itemize}
where $e_i\in\Z_2^5$ is a vector with all zeros except on position $i$. Analogous derivations hold for the other ${\cal M}_i$. For this highly symmetric example, we can follow the next intuitive train of thoughts: $H-H_f$ larger $\Leftarrow$ $H_f$ smaller $\Leftarrow$  $C({\cal G})$ smaller [where ${\cal G}=\sum_{i=1}^{5}{\cal G}_f({\cal M}_i)$] $\Leftarrow$ $t$ smaller $\Leftarrow$ $\mathrm{dim}({\cal G})$ smaller. So we have to choose ${\cal G}_f({\cal M}_i)$ for different $i$ as overlapping as possible to find the highest lower bounds in the LP problem formulation. For this example, there is a one-to-one relationship between $H_f$ and $\mathrm{dim}({\cal G})$:
\[\arr{c||c}{
\mathrm{dim}({\cal G}) & H_f[\mathrm{dim}({\cal G})] \\ \hline \hline
0 & 0\\ \hline
1 & \arr{c}{-[F+\frac{15}{31}(1-F)]\log_2[F+\frac{15}{31}(1-F)]\\-[\frac{16}{31}(1-F)]\log_2[\frac{16}{31}(1-F)]} \\ \hline
2 & \arr{c}{-[F+\frac{7}{31}(1-F)]\log_2[F+\frac{7}{31}(1-F)]\\-3[\frac{8}{31}(1-F)]\log_2[\frac{8}{31}(1-F)]} \\ \hline
3 & \arr{c}{-[F+\frac{3}{31}(1-F)]\log_2[F+\frac{3}{31}(1-F)]\\-7[\frac{4}{31}(1-F)]\log_2[\frac{4}{31}(1-F)]} \\ \hline
4 & \arr{c}{-[F+\frac{1}{31}(1-F)]\log_2[F+\frac{1}{31}(1-F)]\\-15[\frac{2}{31}(1-F)]\log_2[\frac{2}{31}(1-F)]} \\ \hline
5 & -F\log_2 F-31\frac{1-F}{31}\log_2(\frac{1-F}{31})=H}\]
It can then be verified that the yield $\gamma$ equals $1-\sum_{i=1}^5 m({\cal M}_i)=1-5m$ where $m$ is the solution to the following LP problem:
\[\arr{cccl}{
\mbox{minimize} & m & & \\
\mbox{subject to} &  10m &>& H-H_f(0) = H \\
 &  8m &>& H-H_f(1) \\
 &  6m &>& H-H_f(2) \\
 &  4m &>& H-H_f(3) \\
 &  2m &>& H-H_f(4)}\]
Numerical calculation shows that the first inequality is the most binding. Therefore,
\[\gamma=1-\frac{H}{2},\]
which corresponds to our intuition that every measurement yields $n({\cal M}_i)=2$ bits of information. However, this will not be true for a less symmetric mixed state $\rho$. Note that this solution is optimal. Indeed, we cannot gain more than 2 bits per measurement, as $n({\cal M})\leq 2$ for all $\cal M$. We have plotted $\gamma$ as a function of $F$ in Fig.~\ref{figgamma}.

\section{Conclusion}\label{sectionconclusion}

We have presented a breeding protocol that works for any multipartite stabilizer state. Starting with $k$ noisy copies of a stabilizer state and a pool of $(1-\gamma)k$ predistilled pure copies of the same state, the protocol consists of local Clifford operations on noisy and pure copies, followed by local Pauli measurements on the initially pure copies to extract information on the global state. The yield $\gamma$ is calculated as the solution of a linear programming problem. We have illustrated this with a typical example.

\appendix
\section{Construction of orthogonal $A$}\label{apportho}
Before we show how $A$ is constructed, we prove the following theorems.
\begin{theorem}\label{normform}
Any symmetric matrix $W\in\Z_2^{n\times n}$ of rank $r$ can be factorized as follows:
\[W=RDR^T,\]
where $R$ is invertible and
\begin{itemize}
\item[(i)] $D=\mat{cc}{I_{r/2}\otimes\mat{cc}{0&1\\1&0} &  \\  & 0}$ if $W$ has zero diagonal,
\item[(ii)] $D=\mat{cc}{I_r & \\  & 0}$ if $W$ has nonzero diagonal.
\end{itemize}
\end{theorem}

{\bf Proof:}
We prove that if the theorem is true for all $n\leq N$, it also holds for $n=N+1$. Note that the theorem is trivial for zero matrices, as $0=R0R^T$, and matrices of zero dimension.
\begin{itemize}
\item[(i)] Without loss of generality, we may consider (nonzero) $W\in\Z_2^{(N+1)\times(N+1)}$ of the following form:
\[W=\mat{ccc} {0 & 1 & a^T \\ 1 & 0 & b^T \\ a & b & W_2},\]
where $a,b,W_2$ have appropriate dimensions and $W_2$ has zero diagonal.
Indeed, note that identical permutations of rows and columns of $W$ are for free, as they can be absorbed into $R$ as follows:
\[PWP^T=RDR^T \Rightarrow W=(P^TR)D(P^TR)^T.\]
Since $W_2+ab^T+ba^T$ has zero diagonal and is a $(N-1)\times(N-1)$ matrix, we can write:
\[W_2+ab^T+ba^T=R_2D_2R_2^T.\]
It follows that $W=RDR^T=$
\[\mat{ccc}{1 & 0 & 0 \\ 0 & 1 & 0 \\  b & a & R_2}\mat{ccc}{0 & 1 & 0 \\ 1 & 0 & 0 \\  0 & 0 & D_2}\mat{ccc}{1 & 0 & b^T \\ 0 & 1 & a^T \\  0 & 0 & R_2^T}.\]
By construction, $R$ is invertible because $R_2$ is.
\item[(ii)] Again, without loss of generality, we may consider $W$ of the form:
\[W=\mat{cc}{1&a^T\\a&W_2}.\]
We can write:
\[W_2+aa^T=R_2D_2R_2^T,\]
where $D_2$ is either (i) or (ii). It follows that
\[W=RDR^T=\mat{cc}{1 & 0 \\ a & R_2}\mat{cc}{1 & 0 \\ 0 & D_2}\mat{cc}{1 & a^T \\  0 & R_2^T}.\]
If $D_2$ is of the form (i), it can be brought to (ii), by using the identity
\[\mat{ccc}{1&0&0\\0&0&1\\0&1&0}=VV^T,\quad\mbox{with}~V=\mat{ccc}{1&1&1\\1&1&0\\1&0&1}.\]
This ends the proof. \hfill $\square$
\end{itemize}

\begin{corollary}\label{M^TM}
If and only if a given symmetric matrix $W\in\Z_2^{n\times n}$ is not both full rank and zero-diagonal, we can find square $M$ such that $W=M^TM$.
\end{corollary}

{\bf Proof:}
Using Theorem~\ref{normform}, we have $W=RDR^T$. If $D$ is of the form (ii), we take $M$ equal to $R^T$ with the rightmost $n-r$ columns set to zero. If $D$ is of the form (i) and not full rank, we can find $U$ such that $D=U^TU$, by using the identity
\[\mat{ccc}{0&1&0\\1&0&0\\0&0&0}=V^TV,\quad\mbox{with}~V=\mat{ccc}{1&0&0\\1&1&0\\0&1&0}.\]
Then take $M$ equal to $UR^T$ with the rightmost $n-r$ columns set to zero.

Finally, we show that if $W$ is full rank and zero-diagonal, there is no $M$ satisfying $M^TM=W$. An equivalent statement is that there exists no square $M$ such that $M^TM=D=I\otimes\mat{cc}{0&1\\1&0}$. As $x^TDx=0$ for all $x$, $M^TM=D$ implies that $y^Ty=0$ for all $y=Mx$. Consequently, $M$ cannot be full rank. But then $M^TM=D$ cannot be true, as $D$ is full rank. This ends the proof. \hfill $\square$

\begin{theorem}\label{orthogonal}
A matrix $W\in\Z_2^{n\times r}$ can be extended to an orthogonal matrix $A\in\Z_2^{n\times n}$ by adding columns, if and only if
\begin{itemize}
\item $W^TW=I_r$,
\item $e\not\in\col{W}$, where $e\in\Z_2^n$ is the all-ones vector.
\end{itemize}
\end{theorem}

{\bf Proof:}
Define a full rank matrix $Y\in\Z_2^{n\times(n-r)}$ such that $W^TY=0$. By Theorem~\ref{normform}, we can find $R$ and $D$ such that $Y^TY=RDR^T$. For now, we assume that $Y^TY$ is full rank. As $e\not\in\col{W}$, we know that $D=I_{n-r}$. Otherwise $Y^TY$ has a zero diagonal, or $y_i^Ty_i=0$ for all columns $y_i$ of $Y$. Equivalently, we have $y_i^Te=0$ for all $i$, or $Y^Te=0$, which contradicts $e\not\in\col{W}$. Let $Z=YR^{-T}$, then $\mat{cc}{W&Z}$ is orthogonal. Indeed, $Z^TW=R^{-1}Y^TW=0$ and $Z^TZ=R^{-1}Y^TYR^{-T}=D=I$.

This leaves us to proof that $Y^TY$ is full rank. If not, then there exists some $x\neq 0$ that satisfies $Y^TYx=0$. By the definition of $Y$, it follows that $Yx\in\col{W}$ or $Yx=Wz$ for some $z\neq 0$. But then $W^TWz=W^TYx=0$, which contradicts $W^TW=I$. This ends the proof. \hfill $\square$

We now show, for a given full rank $k\times(1-\gamma)k$ matrix Q, how to construct an orthogonal
$\bar{k}\times\bar{k}$ matrix $A$ with lower left part equal to ${Q'}^T$, where $Q'$ is either equal to $Q$ or
equivalent to $Q$ for the protocol. This is the problem adressed in Sec.~\ref{sectionextract}. We perform the
following steps:
\begin{enumerate}
\item find square matrix $M$ such that $M^TM=I+Q^TQ$;
\item create orthogonal $A^T$ from $W=\mat{c}{Q\\M}$ by adding columns.
\end{enumerate}
By Corollary~\ref{M^TM}, step~1 is possible provided that $I+Q^TQ$ is not both full rank and zero diagonal. In that case, this can be solved by adding just one column to $Q$, as the resulting matrix $Q'$ then has an odd number of columns. Consequently, we have one measurement more, but as $k$ is large, this will not influence the yield. By Theorem~\ref{orthogonal}, Step~2 is possible provided that $e\not\in\col{W}$. Let $e\in\col{W}$. Then there exists some $x\neq 0$ that satisfies $Qx=e$ and $Mx=e$. Without loss of generality, we may assume that $x_1=1$. If we add any column (take the second) of $Q$ to the first, yielding $Q'$, and repeat step~1, $e$ will be no longer in $\col{W'}$. This is shown as follows. Let $e\in\col{W'}$, then there exists some $y$ satisfying $Q'y=e$ and $M'y=e$. From the construction of $Q'$ and the fact that $Q$ is full rank, we have $y_i=x_i$ for all $i\neq 2$ and $y_2=1+x_2$. Consequently, $y^Ty\neq x^Tx$. This is contradicted by $x^Tx+y^Ty=x^T(M^TM+Q^TQ)x+y^T({M'}^TM'+{Q'}^TQ')y=0$. Finally, note that $Q'$ is equivalent to $Q$ for the protocol, as the outcomes of the first two measurements in the latter case can be calculated from the corresponding outcomes in the former case.

\begin{acknowledgments}
Research funded by a Ph.D. grant of the Institute for the Promotion of Innovation through Science and Technology in Flanders (IWT-Vlaanderen). Research supported by Research Council KUL: GOA AMBioRICS, CoE EF/05/006 Optimization in Engineering, several PhD/postdoc \& fellow grants; Flemish Government: FWO: PhD/postdoc grants, projects, G.0407.02 (support vector machines), G.0197.02 (power islands), G.0141.03 (identification and cryptography), G.0491.03 (control for intensive care glycemia), G.0120.03 (QIT), G.0452.04 (new quantum algorithms), G.0499.04 (statistics), G.0211.05 (nonlinear), G.0226.06 (cooperative systems and optimization), G.0321.06 (tensors), G.0302.07 (SVM/Kernel), research communities (ICCoS, ANMMM, MLDM); IWT: PhD Grants, McKnow-E, Eureka-Flite2; Belgian Federal Science Policy Office: IUAP P5/22 (`Dynamical Systems and Control: Computation, Identification and Modelling', 2002-2006); EU: ERNSI.
\end{acknowledgments}

\end{document}